\documentclass[aps,floats]{revtex4}
\usepackage{amsmath,amssymb}
\usepackage{graphicx,epsfig}

\begin{document}
\bibliographystyle {plain}

\def\oppropto{\mathop{\propto}} 
\def\opsimeq{\mathop{\simeq}}
\def\opoverderline{\mathop{\overline}}
\def\operarrow{\mathop{\longrightarrow}}
\def\opsim{\mathop{\sim}}

\def\fig#1#2{\includegraphics[height=#1]{#2}}
\def\figx#1#2{\includegraphics[width=#1]{#2}}


\title{ Anderson localization of phonons in dimension $d=1,2,3$ : \\
finite-size properties of the Inverse Participation Ratios of eigenstates } 


 \author{ C\'ecile Monthus and Thomas Garel }
  \affiliation{ Institut de Physique Th\'{e}orique, CNRS and CEA Saclay,
 91191 Gif-sur-Yvette, France}

\begin{abstract}
We study by exact diagonalization the localization properties of phonons in mass-disordered harmonic crystals of dimension $d=1,2,3$. We focus on the behavior of the typical Inverse Participation Ratio $Y_2(\omega,L)$ as a function of the frequency $\omega$ and of the linear length $L$ of the disordered samples. In dimensions $d=1$ and $d=2$, we find that the low-frequency part $\omega \to 0$ of the spectrum satisfies the following finite-size scaling $L Y_2(\omega,L)=F_{d=1}(L^{1/2} \omega)$ in dimension $d=1$ and $L^2 Y_2(\omega,L)=F_{d=2}((\ln L)^{1/2} \omega)$ in dimension $d=2$, with the following conclusions (i) an eigenstate of any fixed frequency $\omega$ becomes localized in the limit $L \to +\infty$ (ii) a given disordered sample of size $L^d$ contains a number $N_{deloc}(L)$ of delocalized states growing as $N_{deloc}(L)\sim L^{1/2}$ in $d=1$ and as $N_{deloc}(L)\sim L^2/(\ln L)$ in $d=2$. In dimension $d=3$, we find a localization-delocalization transition at some finite critical frequency $\omega_c(W)>0$ (that depends on the disorder strength $W$). Our data are compatible with the finite-size scaling  $L^{D(2)} Y_2(\omega,L)=F_{d=3}(L^{1/\nu} (\omega-\omega_c))$ with the values $D(2) \simeq 1.3$ and $\nu \simeq 1.57$ corresponding to the universality class of the localization transition for the Anderson tight-binding electronic model in dimension $d=3$.

\end{abstract}

\maketitle

 \section{ Introduction} 

Since its discovery fifty years ago \cite{anderson,fifty}
Anderson localization has remained a very active field
of research (see for instance the reviews
 \cite{thouless,souillard,bookpastur,Kramer,janssenrevue,markos,mirlinrevue}),
and has been recently realized in experiments 
with atomic matter waves \cite{expe,matterwaves}.
According to the scaling theory \cite{scaltheo}, 
there is no delocalized phase in dimensions $d=1,2$,
whereas there exists a localization/delocalization at finite disorder
in dimension $d>2$. However the notion of Anderson localization
is not limited to quantum electrons models, 
but also applies to classical waves in disordered media \cite{souillard,andersonclassical}
including acoustic waves 
(see for instance \cite{kirkpatrick,laloe,maynard,acoustic} and references therein),
electromagnetic waves (see for instance \cite{john,maret} and references therein),
and hydrodynamical waves \cite{hydro}.
Among the classical disordered models that are expected to display Anderson localization,
the oldest problem is the phonon problem in the presence of random masses coupled by fixed
spring constants, which has been introduced by Dyson \cite{dyson} even before
Anderson's paper \cite{anderson}. After studies concerning the one-dimensional case
(see the review \cite{ishii} and references therein), an analysis of disordered elastic media
via a non-linear sigma-model \cite{john_Sompo} has predicted results similar to 
the scaling theory of Anderson localization \cite{scaltheo} :
all finite-frequency phonons are localized in dimension $d \leq 2$, whereas there exists
a finite critical frequency $\omega_c>0$ in dimension $d>2$ that separates delocalized modes
$\omega<\omega_c$ from localized modes $\omega>\omega_c$. However, in contrast to
electron models where many numerical studies have checked in detail these predictions
and more refined properties like multifractal properties at criticality
 (see the review \cite{mirlinrevue}), the same effort to characterize the statistics of
eigenstates has not been done for the phonon problem.
In particular in dimension $d=3$, the numerical studies we are aware of 
find that almost all states are delocalized, whereas localized states appear only
near band-edges \cite{elliott,akita,sep,Leb_spo}.
In addition, the universality class of the transition does not seem completely clear :
the reported numerical values of the critical exponents are sometimes the same as for
 the Anderson electron transition \cite{elliott}, but are sometimes different 
\cite{akita,sep}.
The aim of the present paper is thus to revisit the problem of phonon localization
in dimension $d=1,2,3$
and to study the properties of the eigenstates Inverse Participation Ratios 
(see definition below in section \ref{model}) which have proven to be very appropriate order parameters of Anderson transitions for electronic models (see the review \cite{mirlinrevue})

The paper is organized as follows. In section \ref{model}, we introduce the phonon model
and the notations for the useful observables. Our numerical exact diagonalization results
are described in the remaining sections. In section \ref{dim1} concerning the one-dimensional case, the finite-size scaling analysis in the low-frequency region is in agreement with
the power-law divergence $\xi(\omega) \propto 1/\omega^2$ of the correlation length
near zero-frequency \cite{ishii,john_Sompo}. In section \ref{dim2} concerning the two-dimensional case, the finite-size scaling analysis in the low-frequency region is in agreement with
the essential-singularity divergence $\ln \xi(\omega) \propto 1/\omega^2$ of the correlation length near zero-frequency \cite{john_Sompo}. Finally in section \ref{dim3} concerning the three-dimensional case, the finite-size scaling analysis around the finite critical frequency is 
compatible with the universality class
of the Anderson transition for the Anderson tight-binding electron model in dimension $d=3$. Our conclusions are summarized in section \ref{sec_conclusion}.

\section{  Model and observables } 

\label{model}

\subsection{ Scalar phonon problem in a crystal of random masses } 

In dimension $d=1,2,3$, we consider $L^d$ random masses $m_{\vec r}$ whose positions
$\vec r=(n_1,..n_d)$ at rest form an hypercubic lattice ($n_i=1,2,..,L)$. 
These masses are coupled by spring constants
$K_{\vec r,\vec r \ '}=1$ if $\vec r$ and $\vec r \ '$ are neighbors on the hypercubic lattice, so that each mass in the bulk has $(2d)$ neighbors.
We consider the following 
 harmonic Hamiltonian for the scalar displacements $u_{\vec r}(t)$
\begin{eqnarray}
H = \sum_{\vec r} \frac{ m_{\vec r}}{2} \dot{u}_{\vec r}^2 
+ \sum_{\vec r, \vec r \ '} \frac{K_{\vec r,\vec r \ '}}{2} (u_{\vec r}-u_{\vec r \ '})^2 
\label{Hphonon}
\end{eqnarray}
The scalar assumption is very standard to simplify the analysis
\cite{Leb_spo} and means physically that longitudinal and transverse vibrations are decoupled. 
Equivalently, the model can be defined by the equations of motion 
\begin{eqnarray}
 m_{\vec r} \ddot{u}_{\vec r} = 
- \sum_{\vec r \ '} K_{\vec r,\vec r \ '} 
\left(u_{\vec r}-u_{\vec r \ '}\right)
\label{motion}
\end{eqnarray}
To avoid the free-motion of the center-of-mass of the system, we have chosen to
consider the fixed boundary conditions $u=0$ on the lines $n_i=0$ and $n_i=L+1$
surrounding the hypercube.

Finally, to avoid the peculiarities introduced by a binary distribution of the disorder
 \cite{dean,Leb_spo},
we have chosen to consider
 the continuous flat distribution for the random masses $m_{\vec r}$.
\begin{eqnarray}
P(m) = \frac{1}{W } \theta \left( 1 \leq m \leq 1+W \right)
\label{pm}
\end{eqnarray}
so that $W$ represents the disorder strength. 
The numerical results presented in this paper correspond to the two cases
$W=1$ and $W=20$.

\subsection{ Eigenmodes analysis }

Since the equations of motion of Eq. \ref{motion} are linear, 
the dynamics can be analyzed via the eigenmodes of oscillations in $e^{i \omega t}$ :
the eigenvalues $\omega_p^2$ and 
the associated eigenmodes $a_{p}(\vec r)$ satisfy
\begin{eqnarray}
 m_{\vec r} \omega_p^2 a_{p}(\vec r) = 
 \sum_{\vec r \ '} K_{\vec r,\vec r \ '} 
\left(a_p({\vec r})-a_p({\vec r \ '}) \right)
\label{eigen}
\end{eqnarray}
It is more convenient to
perform the similarity transformation \cite{dyson}
\begin{eqnarray}
 a_{p}(\vec r) = \frac{\psi_{p}(\vec r)}{\sqrt{m_{\vec r} }}
\label{similarity}
\end{eqnarray}
to reduce the problem to the diagonalization of a symmetric operator
\begin{eqnarray}
 \omega_p^2   \psi_{p}(\vec r)
= \left[  
\frac{ \sum_{\vec r \ '} K_{\vec r,\vec r \ '} } { m_{\vec r}  } \right] \psi_{p}(\vec r)
- 
\sum_{\vec r \ '} \frac{ K_{\vec r,\vec r \ '} } {\sqrt{m_{\vec r} m_{\vec r \ '} }} \  \psi_{p}(\vec r \ ')
\label{eigensym}
\end{eqnarray}
As stressed in \cite{Leb_spo}, this form is analog to an Anderson tight-binding model with on-site energies $\epsilon(\vec r)= \left[  \sum_{\vec r \ '}
\frac{ K_{\vec r,\vec r \ '} } { m_{\vec r}  } \right]$ and 
hoppings $\frac{ K_{\vec r,\vec r \ '} } {\sqrt{m_{\vec r} m_{\vec r \ '} }}$,
but as a consequence of correlations through the random masses, different physical properties can occur. In particular, the eigenvalues are positive $E_p=\omega_p^2 \geq 0$ in the phonon problem, whereas $E=0$ is the center of the band in usual Anderson tight-binding models. It is convenient to work 
with the orthogonal basis $\phi_p$ of eigenvectors of Eq. \ref{eigensym}
normalized to
\begin{eqnarray}
<\phi_p \vert \phi_p > = \sum_{\vec r} \phi_p^2(\vec r) =1
\label{normaphi}
\end{eqnarray}
This means that the phonons eigenmodes of Eq. \ref{similarity} are normalized according to
\begin{eqnarray}
1 = \sum_{\vec r} m_{\vec r} a_p^2(\vec r) 
\label{normaa}
\end{eqnarray}

\subsection{ Inverse Participation Ratios (I.P.R.) }

To characterize the localization properties of the phonon eigenmodes $a_{p}(\vec r)$
introduced above in Eq. \ref{eigen}, we consider 
 the Inverse Participation Ratio (I.P.R.)
\begin{eqnarray}
Y_2(\omega_p,L) \equiv \frac{ \sum_{\vec r} a_{p}^{4}(\vec r)}{ \left( \sum_{\vec r} a_{p}^2(\vec r) \right)^2}
\label{ipr}
\end{eqnarray}
that represents an order parameter 
for Anderson localization transition \cite{mirlinrevue} :
at a given frequency $\omega$, localized eigenstates correspond to a finite value 
 in the limit $L \to +\infty$
\begin{eqnarray}
Y_2^{loc}(\omega,L) \oppropto_{L \to +\infty} Y_2(\omega,\infty) >0
\label{iprloc}
\end{eqnarray}
 whereas delocalized states
correspond to the following power-law decay
\begin{eqnarray}
Y_2^{deloc}(\omega,L) \oppropto_{L \to +\infty} \frac{1}{L^d}
\label{iprdeloc}
\end{eqnarray}

Note that for phonons, the standard definition of the I.P.R. $Y_2$
is Eq. \ref{ipr} in terms of the modes $a_{p}(\vec r)$ \cite{Leb_spo}
that are normalized with Eq. \ref{normaa},
whereas in electronic tight-binding models, the I.P.R. are defined in terms
of the orthogonal basis $\phi_p$ \cite{mirlinrevue}
normalized with Eq. \ref{normaphi}.
Since the random masses appearing in the normalization of Eq. \ref{normaa}
remain bounded (see the distribution of Eq. \ref{pm}), we believe that the choice
of the $a_p$ or of the $\phi_p$ to compute the I.P.R. should 
of course affect their precise numerical values, but 
should not change their scaling properties with the system size $L$.
In particular, the localization and delocalization criterions of Eqs \ref{iprloc}
and \ref{iprdeloc} should give the same results for the two definitions.
In the following, all numerical results correspond to the definition of Eq. 
\ref{ipr}.

\subsection{ Average over the disordered samples of a given size }

In practice, for each size $L$ in dimension $d$, 
we generate a certain number $n_S(L)$
of disordered samples containing $L^d$ random masses. 
The exact diagonalization of each sample $\alpha$
(via the standard NAG diagonalization routine F02FAF
that computes all the eigenvalues and all the eigenvectors of a real symmetric matrix)
 yields the $L^d$ eigenmodes $a_{p}^{\alpha}$
that are ordered by their frequency $\omega_p^{(\alpha)}$ in ascending order
$0 < \omega_1^{(\alpha)}<\omega_2^{(\alpha)}<...<\omega_{L^d}^{(\alpha)}$.
For each index $p=1,..,L^d$, we have computed the typical frequency
\begin{eqnarray}
\omega_p^{typ}(L) \equiv e^{ \overline {\ln \omega_p^{(\alpha)} }}
\label{omegaityp}
\end{eqnarray}
and the corresponding typical I.P.R.
\begin{eqnarray}
Y_p^{typ} (L)\equiv e^{ \overline {\ln Y_p^{(\alpha)} (L) }}
\label{y2ityp}
\end{eqnarray}
where $\overline{A}$ denotes the average of the observable $A$
over the disordered samples $(\alpha)$.
The integrated density of states is then obtained as
\begin{eqnarray}
N(\omega) = \frac{1}{L^d} \sum_{p=1}^{L^d} \theta( \omega_p^{typ}(L) \leq \omega )
\label{integrateddos}
\end{eqnarray}
with the boundaries values $N(0)=0$ and $N(+\infty)=1$.
The parametric plot $(\omega_p^{typ}(L),Y_p^{typ} (L))$ with $p=1,2,..L^d$
allows to obtain the behavior of the I.P.R. $Y_2^{typ}(\omega,L)$ as a function
of the frequency $\omega$ and of the length $L$.

\section{ Localization properties of phonons in dimension $d=1$ }

\label{dim1}

In this section, we present our numerical results obtained in dimension $d=1$.
for the following sizes $L$ and the corresponding 
number $n_s(L)$ of disordered samples
\begin{eqnarray}
L && = 10^2, 2.10^2, 5.10^2, 10^3, 2.10^3, 3.10^3, 4.10^3, 5.10^3
\nonumber \\
n_s(L) && = 2.10^7, 47.10^5 , 43.10^4, 4.10^4, 3.10^3, 10^3, 350, 150
\label{nume1d}
\end{eqnarray}

\subsection{ Density of states  }

\begin{figure}[htbp]
 \includegraphics[height=6cm]{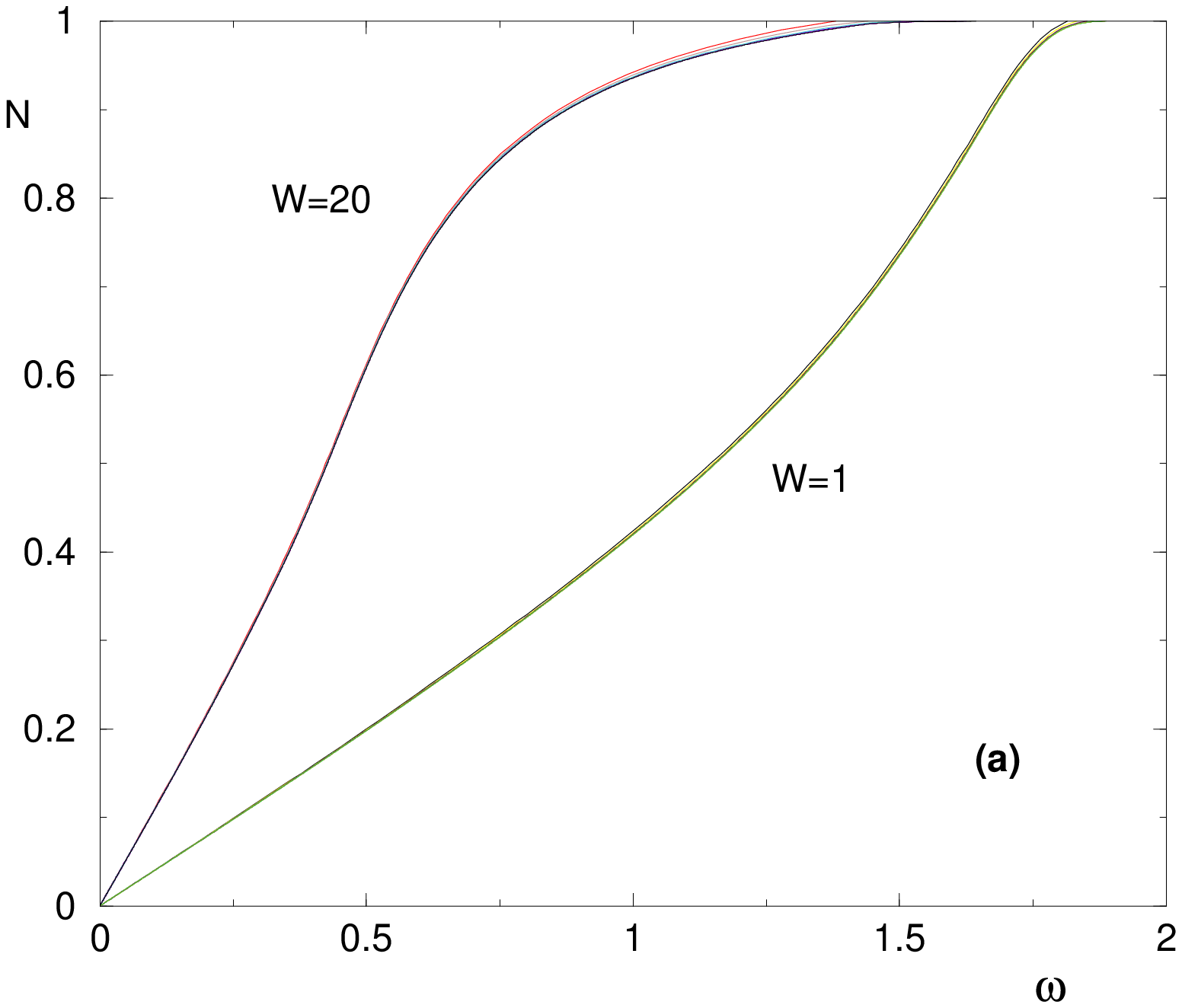}
\vspace{1cm}
 \includegraphics[height=6cm]{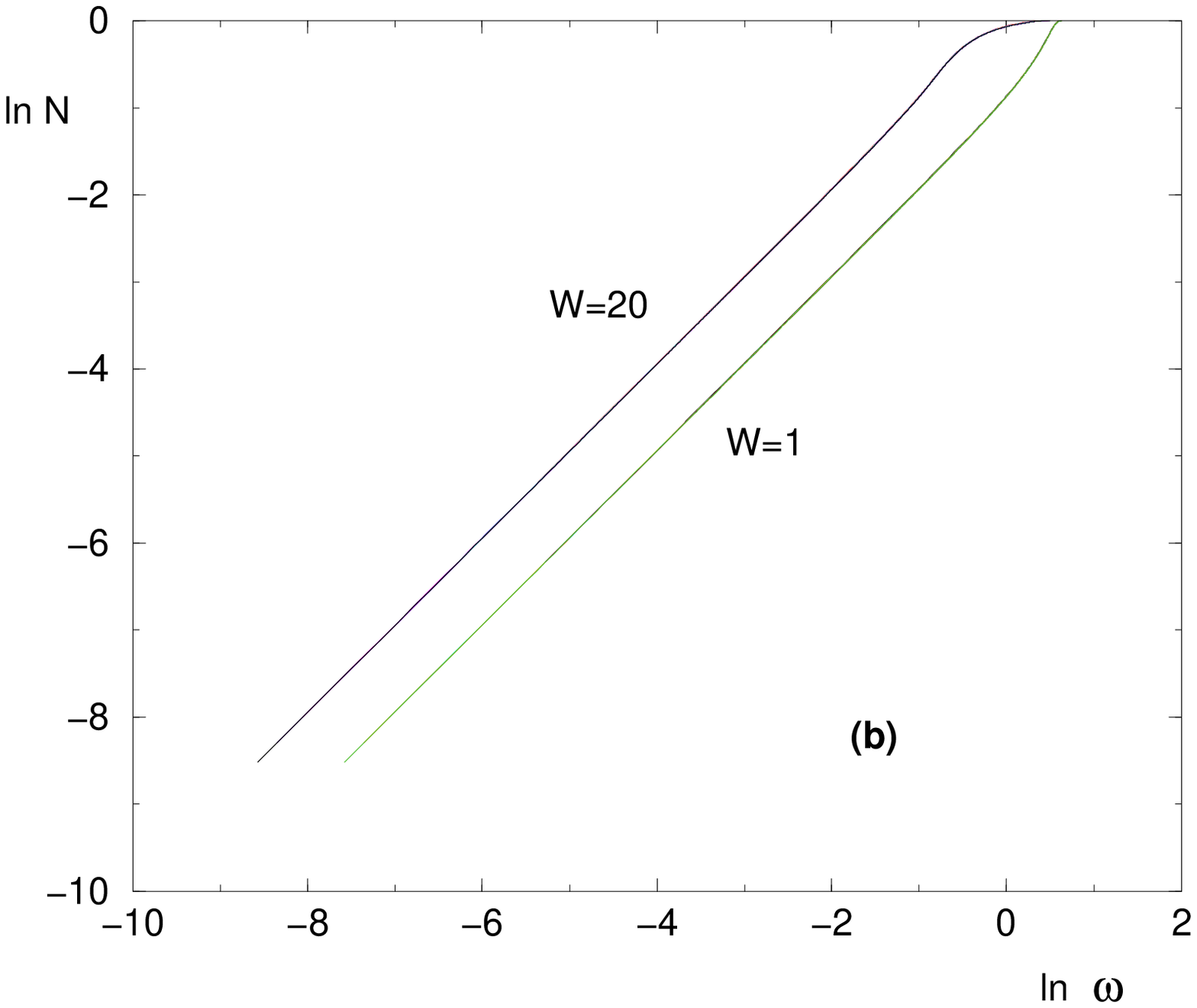}
\caption{(Color on line)  Integrated density of states $N(\omega)$ of Eq. \ref{integrateddos} for phonons in $d=1$
(a) $N(\omega)$ for various sizes $100 \leq L \leq 5000$ and two disorder strengths $W=1$ and $W=20$
(b) same data in log-log scales to display the low-frequency behavior of Eq. 
\ref{integrateddos1d} 
}
\label{figdos1d}
\end{figure}

We show on Fig. \ref{figdos1d} (a) the integrated density of states $N(\omega)$ of 
Eq. \ref{integrateddos} for two disorder strengths $W=1$ and $W=20$.
As shown in log-log scale on Fig. \ref{figdos1d} (b), we find the linear behavior
already present in the pure case
\begin{eqnarray}
N(\omega) \opsimeq_{\omega \to 0} C(W) \omega
\label{integrateddos1d}
\end{eqnarray}
and the disorder strength $W$ is only present in the numerical prefactor $C(W)$.
We have also checked that the lowest frequency mode scales as $\omega_1(L) \propto 1/L$.

\subsection{  Typical Inverse Participation Ratio  $Y_2^{typ}(\omega,L)$}

\begin{figure}[htbp]
 \includegraphics[height=6cm]{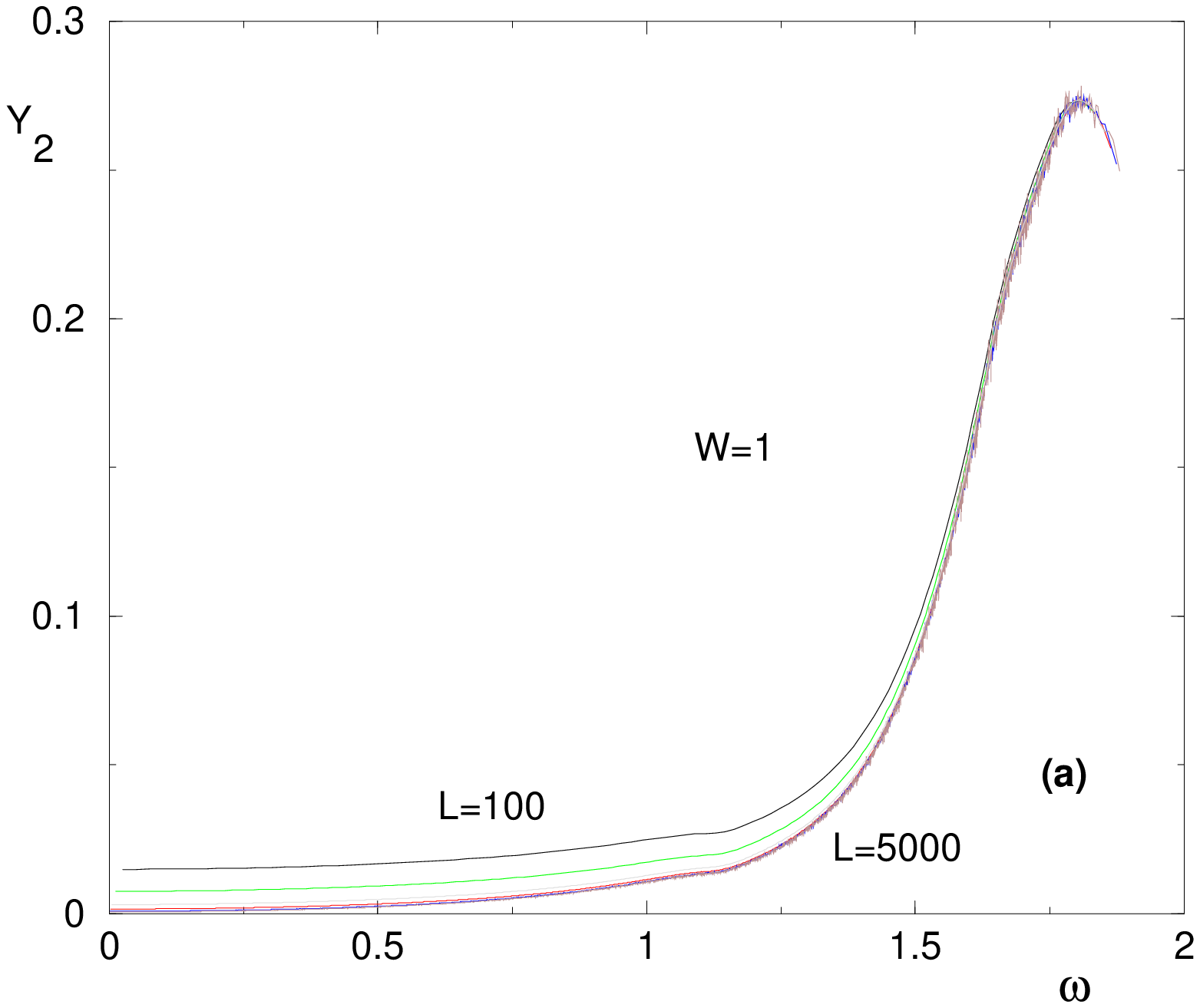}
\vspace{1cm}
 \includegraphics[height=6cm]{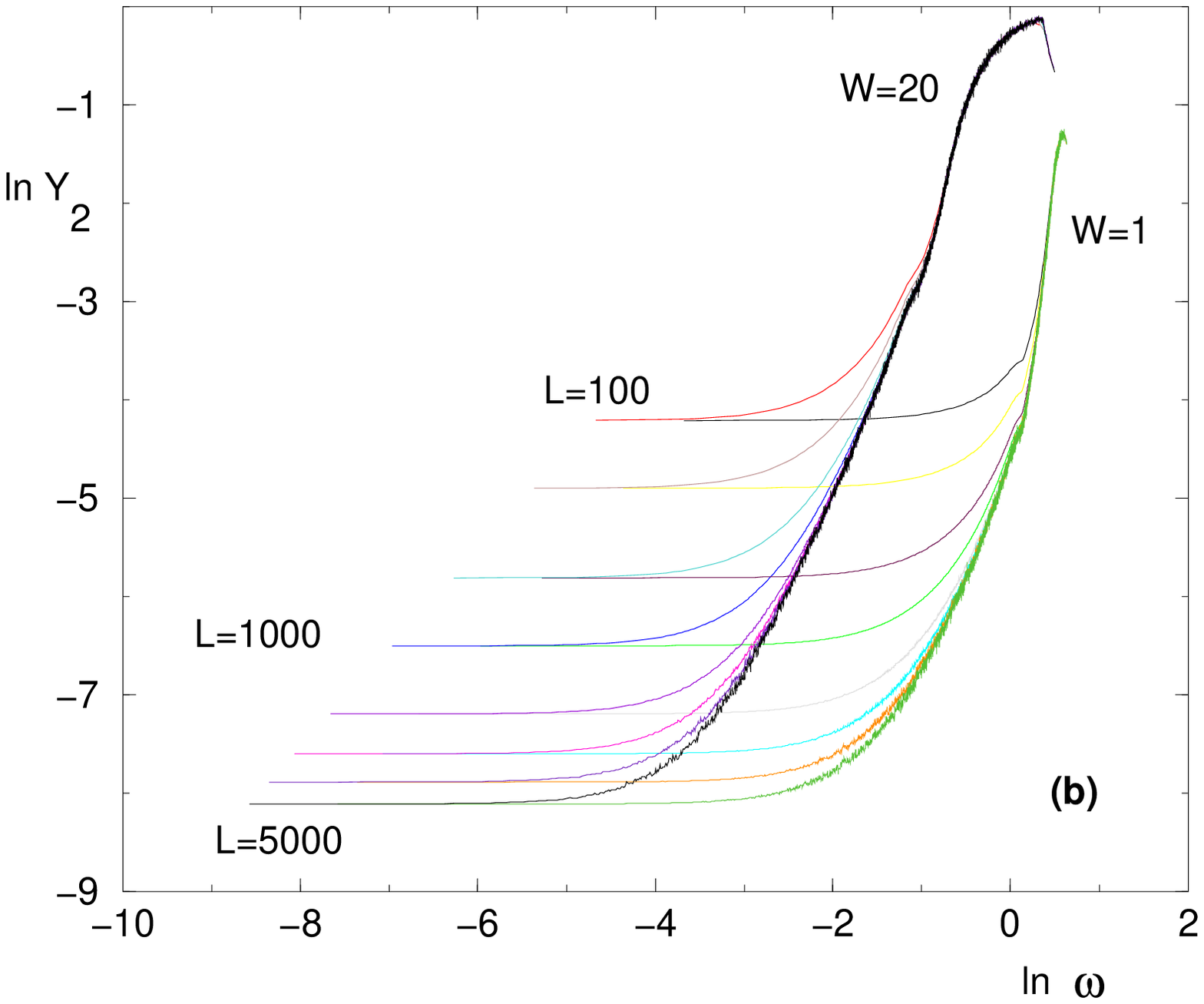}
\caption{(Color on line) Typical Inverse Participation Ratio $Y_2(\omega,L)$ as a function of the frequency 
$\omega$ in $d=1$ for various sizes $100 \leq L \leq 5000$
(a) $ Y_2^{typ}(\omega,L)$ as a function of $ \omega$ for the disorder strength $W=1$ 
(b) $\ln Y_2^{typ}(\omega,L)$ as a function of $\ln \omega$ 
for two disorder strengths $W=1$ and $W=20$ }
\label{figomipp1d}
\end{figure}

To analyse the localization properties of eigenstates, we show on Fig. \ref{figomipp1d}
the typical Inverse Participation Ratio $Y_2^{typ}(\omega,L)$ of Eq. \ref{y2ityp}
as a function of the frequency $\omega$ for various sizes $L$.
In the high-frequency domain where the data of all sizes collapse,
the eigenstates are localized. In the low-frequency domain where all sizes give different results, as shown more clearly in log-log scale on Fig. \ref{figomipp1d} (b),
eigenstates are delocalized on the whole disordered sample.
We find moreover that the data for the very different disorder strengths $W=1$ and $W=20$ merge in the low-frequency region for each size $L$ : this means that the lowest frequencies eigenstates are delocalized in the same way independently of the disorder strength
$W$.

\subsection{ Finite-size scaling analysis of the low-frequency modes }

\begin{figure}[htbp]
 \includegraphics[height=6cm]{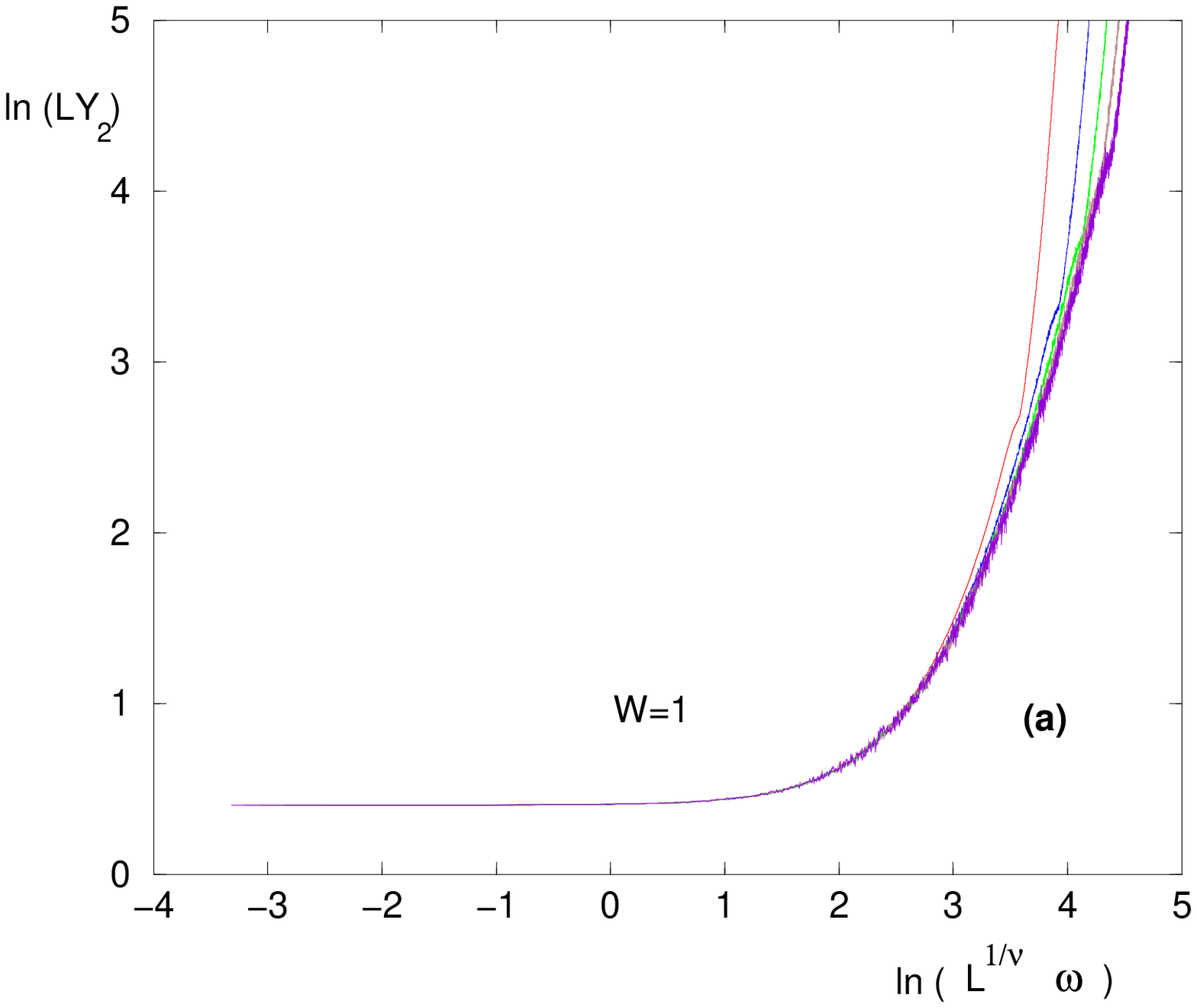}
\vspace{1cm}
 \includegraphics[height=6cm]{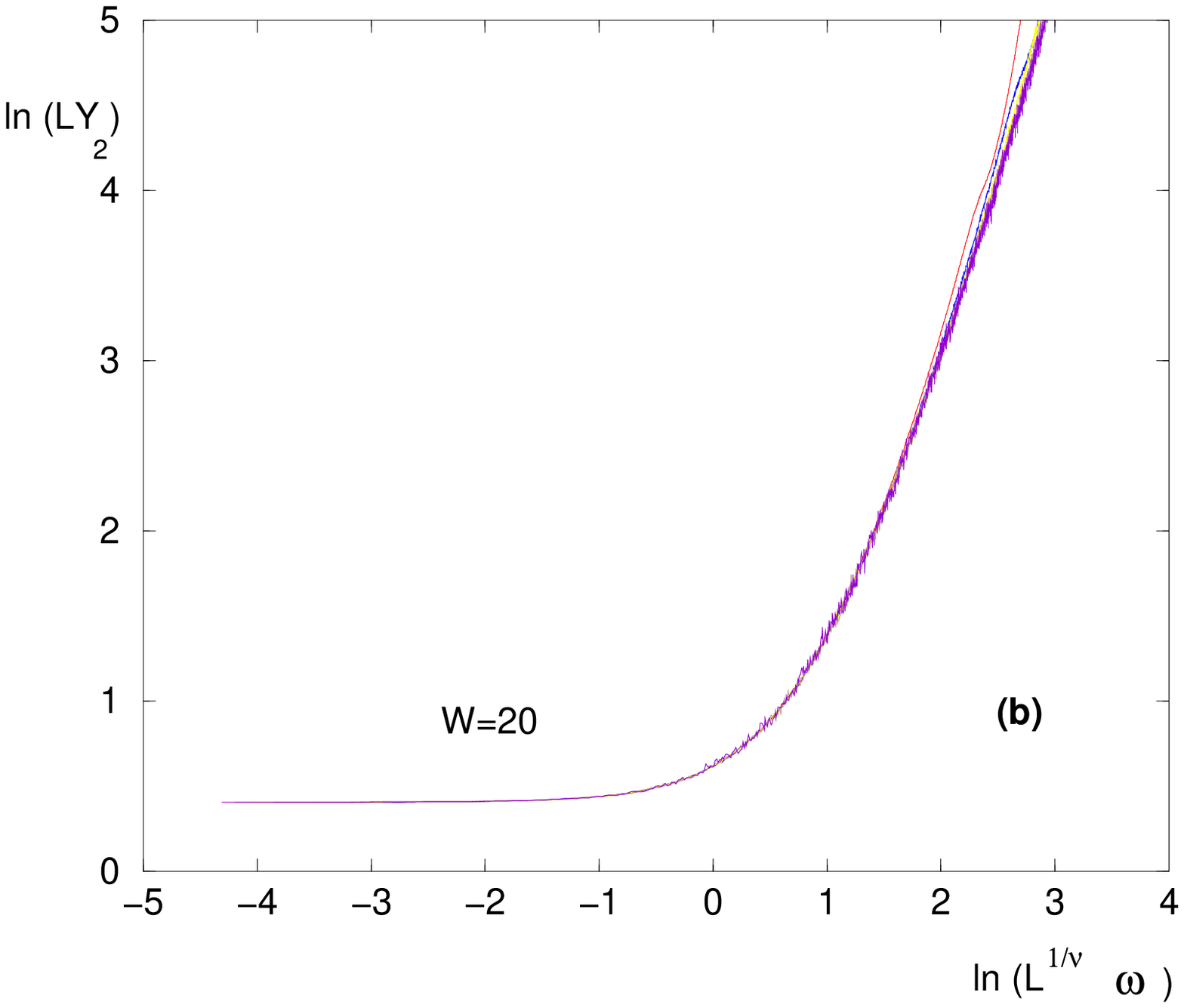}
\caption{(Color on line)  Finite-size scaling analysis of the 
typical I.P.R. $Y_2^{typ}(\omega,L)$ 
of the low-frequency eigenmodes in $d=1$ 
according to Eq. \ref{fssy2d=1} : the rescaled variable 
$y= L Y_2^{typ} (\omega,L)$ is plotted as a function of
the reduced variable $x=L^{1/\nu} \omega $ with the value $\nu=2$
(a) data collapse in log-log scale for $W=1$
(b) data collapse in log-log scale for $W=20$ }
\label{figfss1d}
\end{figure}

We show on Fig. \ref{figfss1d} that our data are compatible with the following 
finite-size scaling for the low-energy modes
\begin{eqnarray}
Y^{typ}_2 (\omega,L) \simeq \frac{1}{L} F_{d=1} \left( L^{1/\nu} \omega \right) \ \ {\rm with }
\ \ \nu=2
\label{fssy2d=1}
\end{eqnarray}
This means that to each frequency $\omega$, one can associate a correlation length
$\xi(\omega)$ diverging as the power-law
\begin{eqnarray}
\xi(\omega) \oppropto_{\omega \to 0} \frac{1}{\omega^{2}}
\label{xid=1}
\end{eqnarray}
in agreement with \cite{ishii,john_Sompo}.
An eigenmode of fixed frequency $\omega$ will be delocalized on samples
 of small lengths $L \ll \xi(\omega)$ with an I.P.R. of order 
 \begin{eqnarray}
Y^{typ}_2 (\omega,L) \opsimeq_{L \ll \xi(\omega) }
 \frac{1}{L} F_{d=1} \left( 0 \right)
\label{fssy2d=1deloc}
\end{eqnarray}
but will be localized on samples of large lengths $L \gg \xi(\omega)$
 with an I.P.R. of order (using $F_{d=1}(x) \propto x^{\nu}$ at large $x$)
 \begin{eqnarray}
Y^{typ}_2 (\omega,L) \opsimeq_{L \gg \xi(\omega) }  \omega^{2} = \frac{1}{\xi(\omega)}
\label{fssy2d=1loc}
\end{eqnarray}
The conclusion is thus that for any fixed frequency $\omega$,
the corresponding eigenmodes will become localized in
the limit $L \to +\infty$  \cite{ishii,john_Sompo}.
However if one is interested into the set of eigenstates
 of a sample of a given size $L$,
the conclusion is that frequencies $\omega \ge L^{-1/2}$ correspond
to localized eigenmodes,
whereas a certain number ${\cal N}_{deloc}(L)$ of eigenmodes corresponding
to frequencies $\omega \le L^{-1/2}$ are delocalized. 
From the linear behavior in $\omega$
of the integrated density of states of Eq. \ref{integrateddos1d}, one obtains that 
the fraction of delocalized states scales as the pseudo-critical value 
$\omega^*(L)\sim L^{-1/2} $
 \begin{eqnarray}
\frac{{\cal N}_{deloc}(L)}{L } \propto \omega^*(L) \sim L^{-1/2}
\label{fracndeloc1d}
\end{eqnarray}
So the number of delocalized eigenstates in a sample of size $L$ grows as 
 \begin{eqnarray}
{\cal N}_{deloc}(L) \oppropto_{L \to +\infty}  L^{1/2}
\label{ndeloc1d}
\end{eqnarray}
in agreement with the interpretation given in \cite{ishii}.
Note that this property of phonons in dimension $d=1$ is very different from
the Anderson electronic problem, where
 the whole set of eigenstates of a given sample 
become localized at large sizes.
Physically, this difference is essential if one considers the dynamical
properties, since the dynamics can be expanded on the basis of eigenmodes :
in the Anderson electronic problem, the localization of the whole set
of eigenfunctions imply the exponential localization for the dynamical
problem starting from any localized initial condition, whereas in the phonon
case, the presence of these low-frequency delocalized modes for any size
prevents the exponential localization in the dynamics.

\section{ Localization properties of phonons in dimension $d=2$ } 

\label{dim2}

In this section, we present our numerical results obtained in dimension $d=2$.
for the following sizes $L$ and the corresponding 
number $n_s(L)$ of disordered samples
\begin{eqnarray}
L && = 10, 20, 30, 40, 50, 60, 70, 80
\nonumber \\
n_s(L) && = 2.10^7,7.10^5, 77.10^3, 7500, 1800, 500, 250, 200
\label{nume2d}
\end{eqnarray}

\subsection{ Density of states  }

\begin{figure}[htbp]
 \includegraphics[height=6cm]{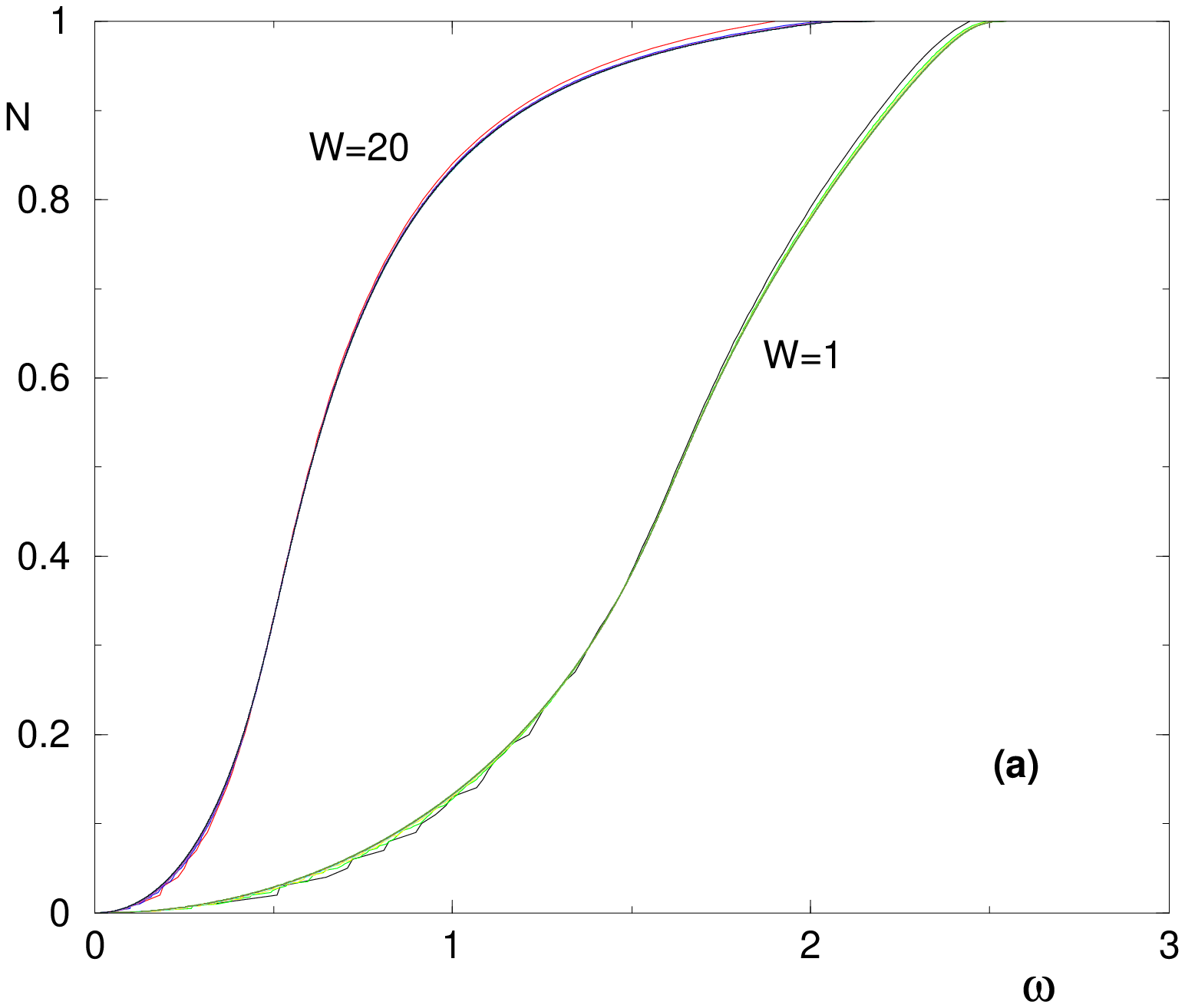}
\vspace{1cm}
 \includegraphics[height=6cm]{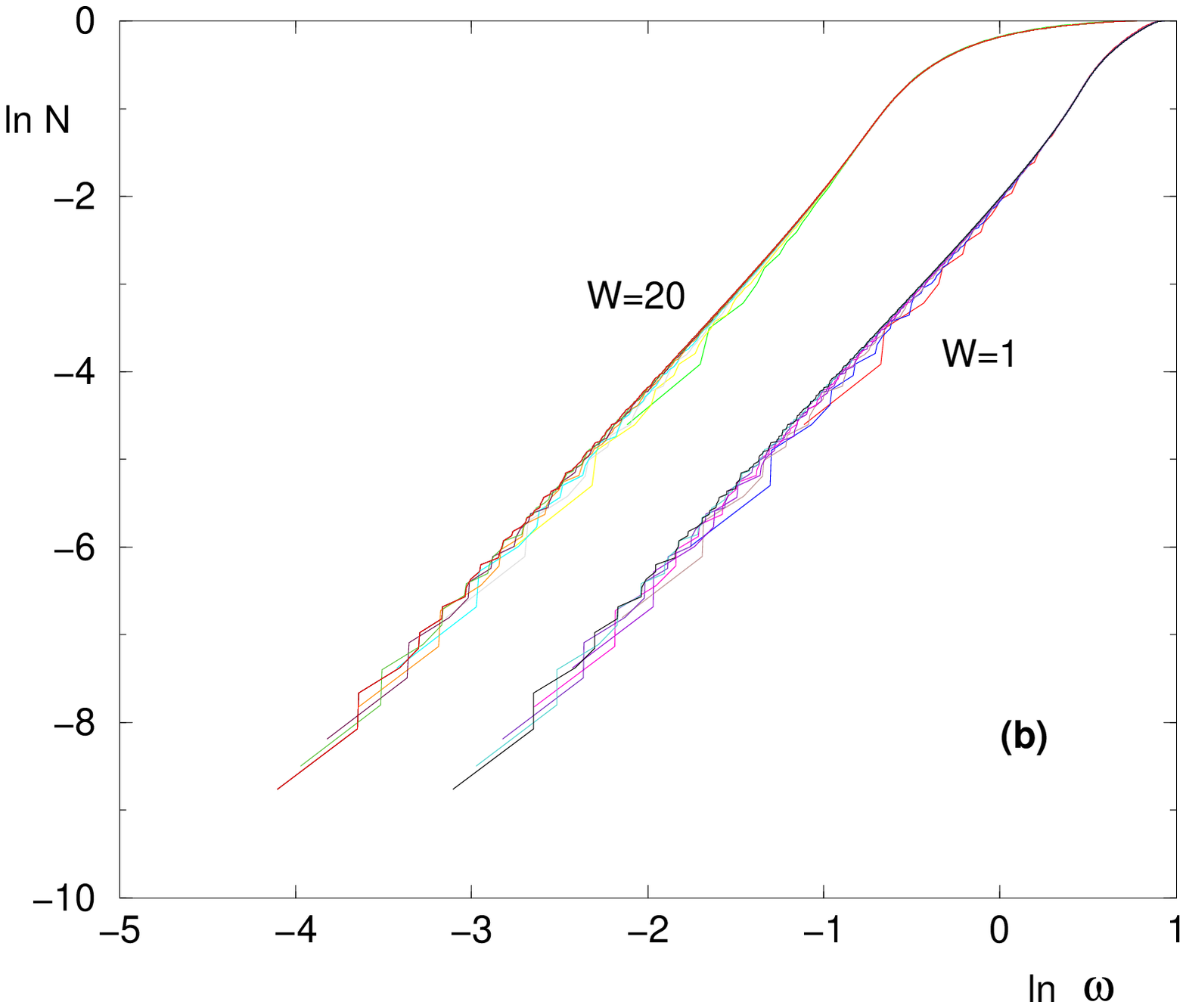}
\caption{(Color on line)  Integrated density of states $N(\omega)$
for phonons in $d=2$
(a) $N(\omega)$ for various sizes $10 \leq L \leq 80$ and two disorder strengths $W=1$ and $W=20$
(b) same data in log-log scales to display the low-frequency behavior 
 of Eq. \ref{integrateddos2d}  }
\label{figdos2d}
\end{figure}

We show on Fig. \ref{figdos2d} (a) the integrated density of states $N(\omega)$ of 
Eq. \ref{integrateddos} for two disorder strengths $W=1$ and $W=20$.
As shown in log-log scale on Fig. \ref{figdos2d} (b), we find the same behavior
as in the pure case
\begin{eqnarray}
N(\omega) \oppropto_{\omega \to 0}  \omega^2
\label{integrateddos2d}
\end{eqnarray}
and the disorder strength $W$ is only present in the numerical prefactor.
We have also checked that the lowest frequency mode scales as $\omega_1(L) \propto 1/L$.

\subsection{  Typical Inverse Participation Ratio $Y_2^{typ}(\omega,L)$ }

\begin{figure}[htbp]
 \includegraphics[height=6cm]{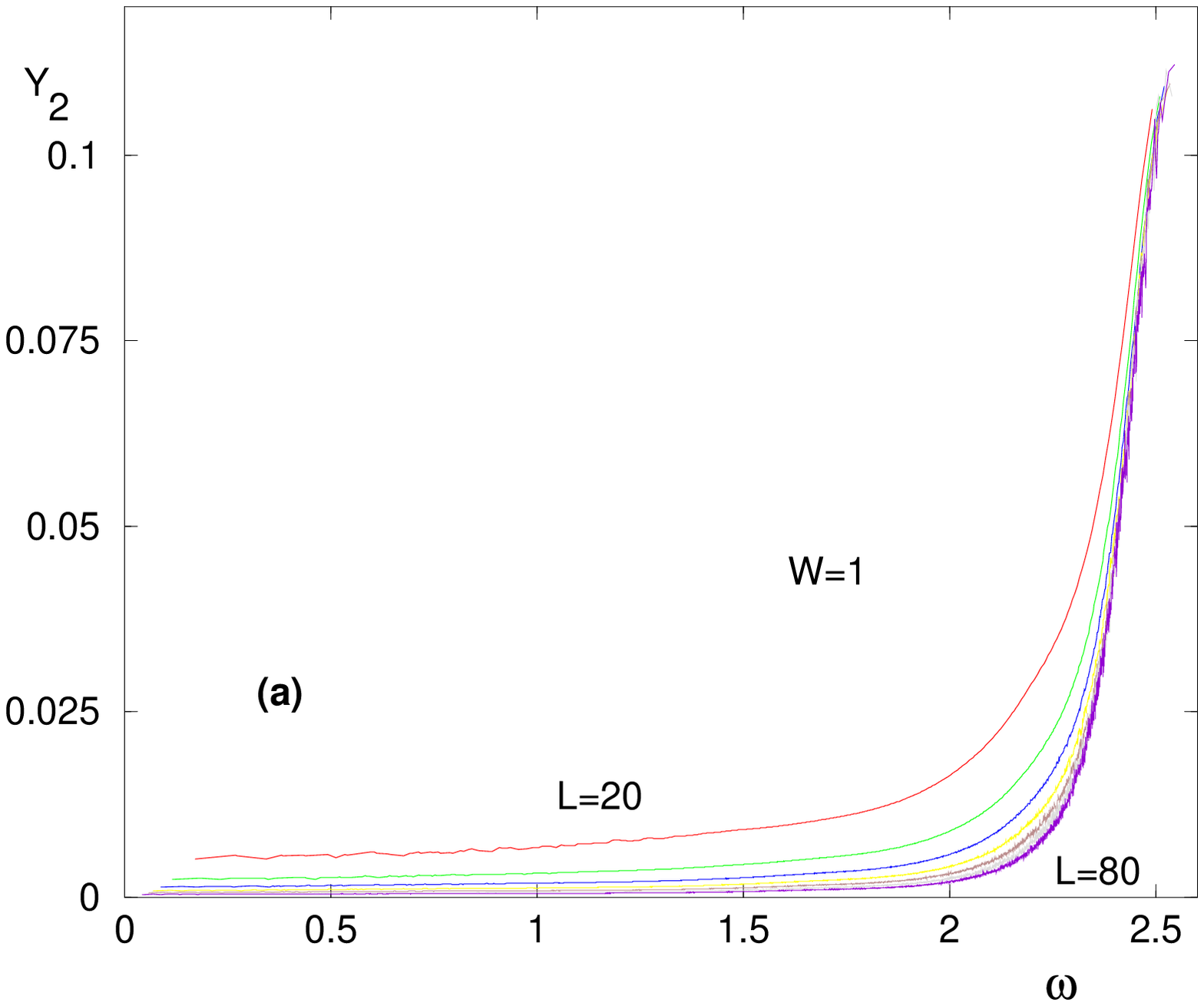}
\vspace{1cm}
 \includegraphics[height=6cm]{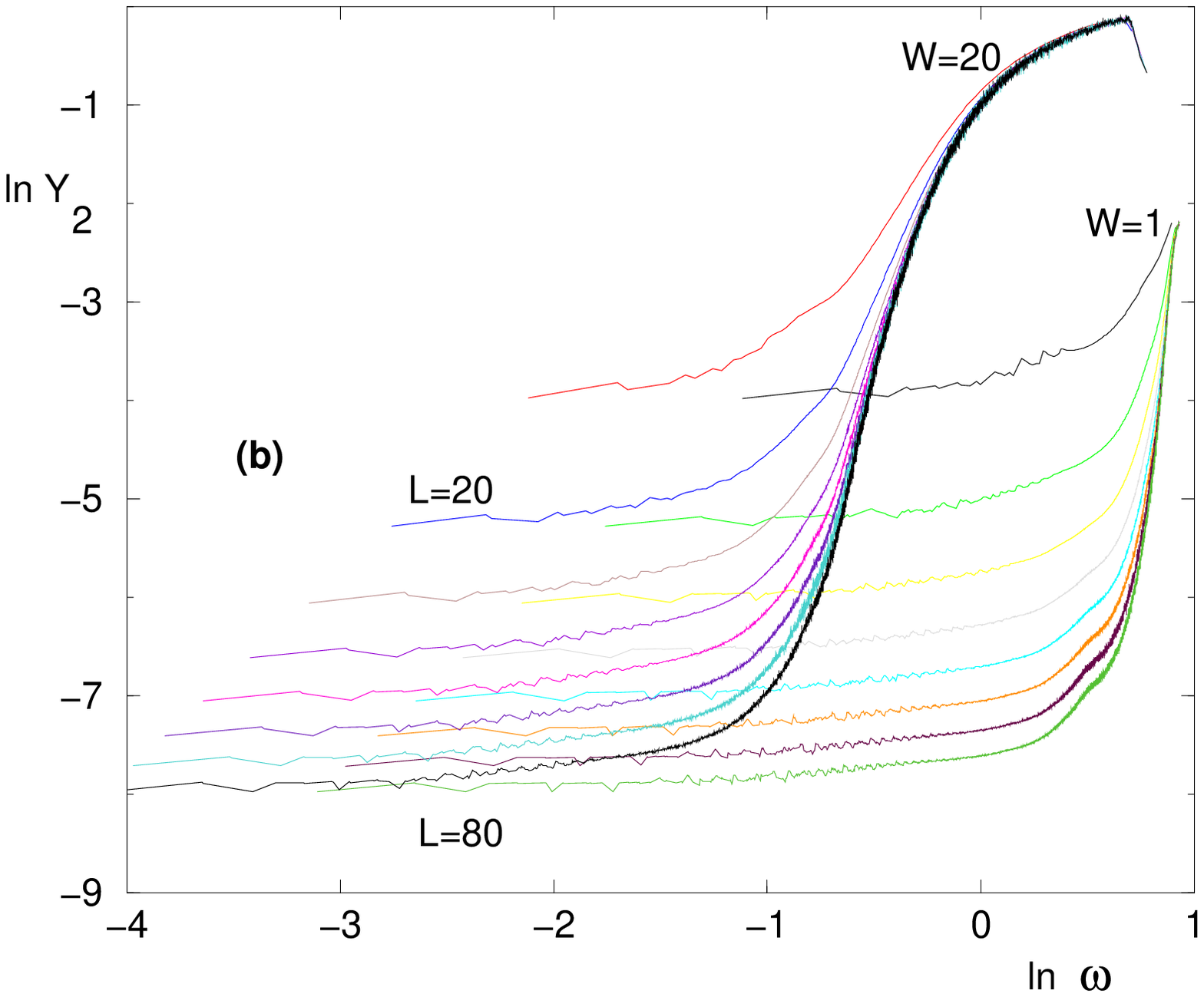}
\caption{(Color on line)  Typical IPR $Y_2^{typ}(\omega,L)$ as a function of the frequency 
$\omega$ in $d=2$ for
all eigenmodes of samples of sizes $20 \leq L \leq 80$
(a) $ Y_2^{typ}(\omega,L)$ as a function of $ \omega$ for $W=1$ 
(b) $\ln Y_2^{typ}(\omega,L)$ as a function of $\ln \omega$ 
for two disorder strengths $W=1$ and $W=20$ }
\label{figomipp2d}
\end{figure}

We show on Fig. \ref{figomipp2d}
the typical Inverse Participation Ratio $Y_2^{typ}(\omega,L)$ of Eq. \ref{y2ityp}
as a function of the frequency $\omega$ for various sizes $L$.
In the high-frequency domain where the data of all sizes collapse,
the eigenstates are localized. In the low-frequency domain where all sizes give different results, as shown more clearly in log-log scale on Fig. \ref{figomipp2d} (b),
eigenstates are delocalized on the whole disordered sample.
As in dimension $d=1$, we find moreover that the data for the two disorder strengths $W=1$ and $W=20$ merge in the low-frequency region for each size $L$ : this means that the lowest frequencies eigenstates are delocalized in the same way independently of the disorder strength.

\subsection{ Finite-size scaling analysis of the low-frequency modes }

\begin{figure}[htbp]
 \includegraphics[height=6cm]{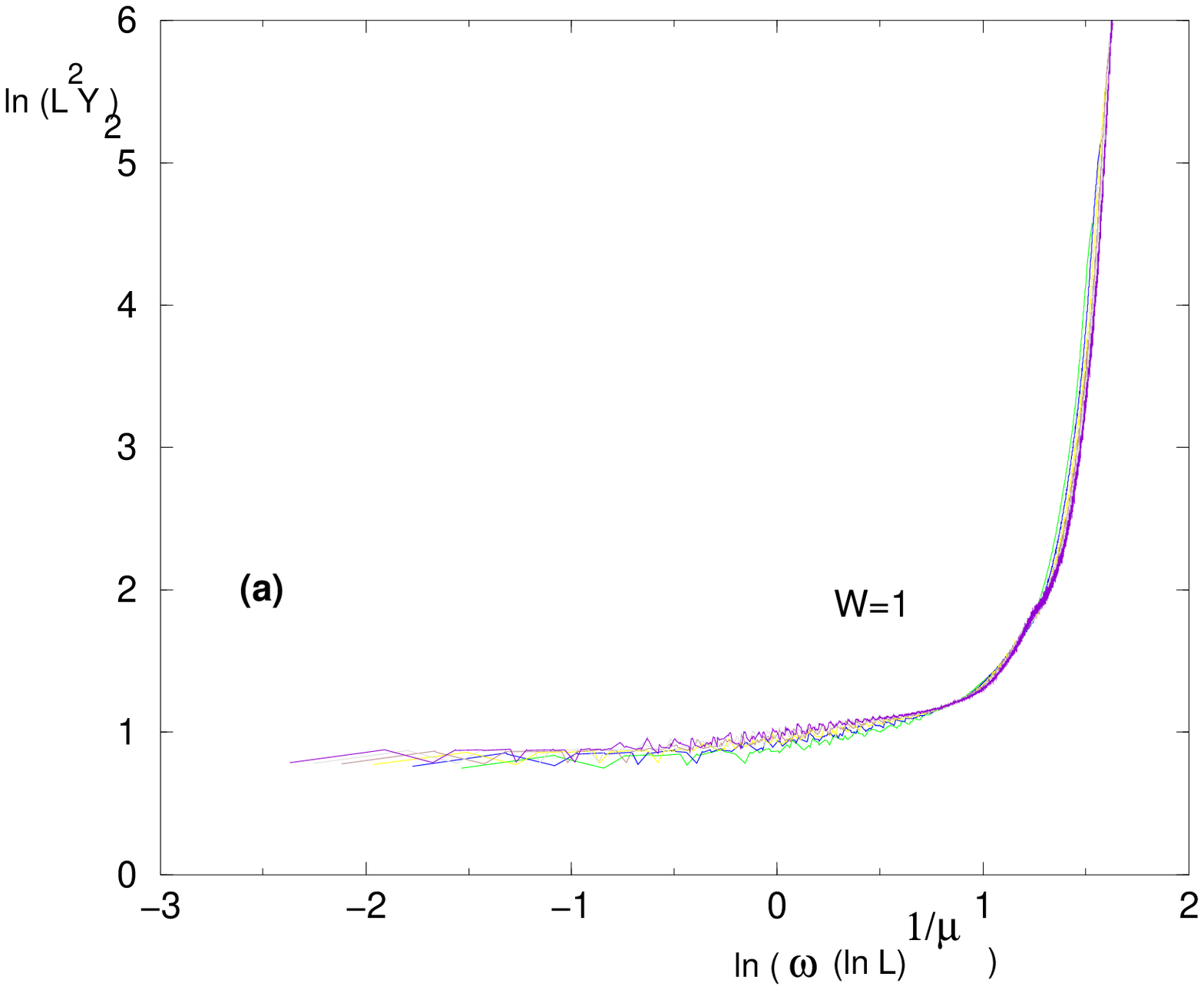}
\vspace{1cm}
 \includegraphics[height=6cm]{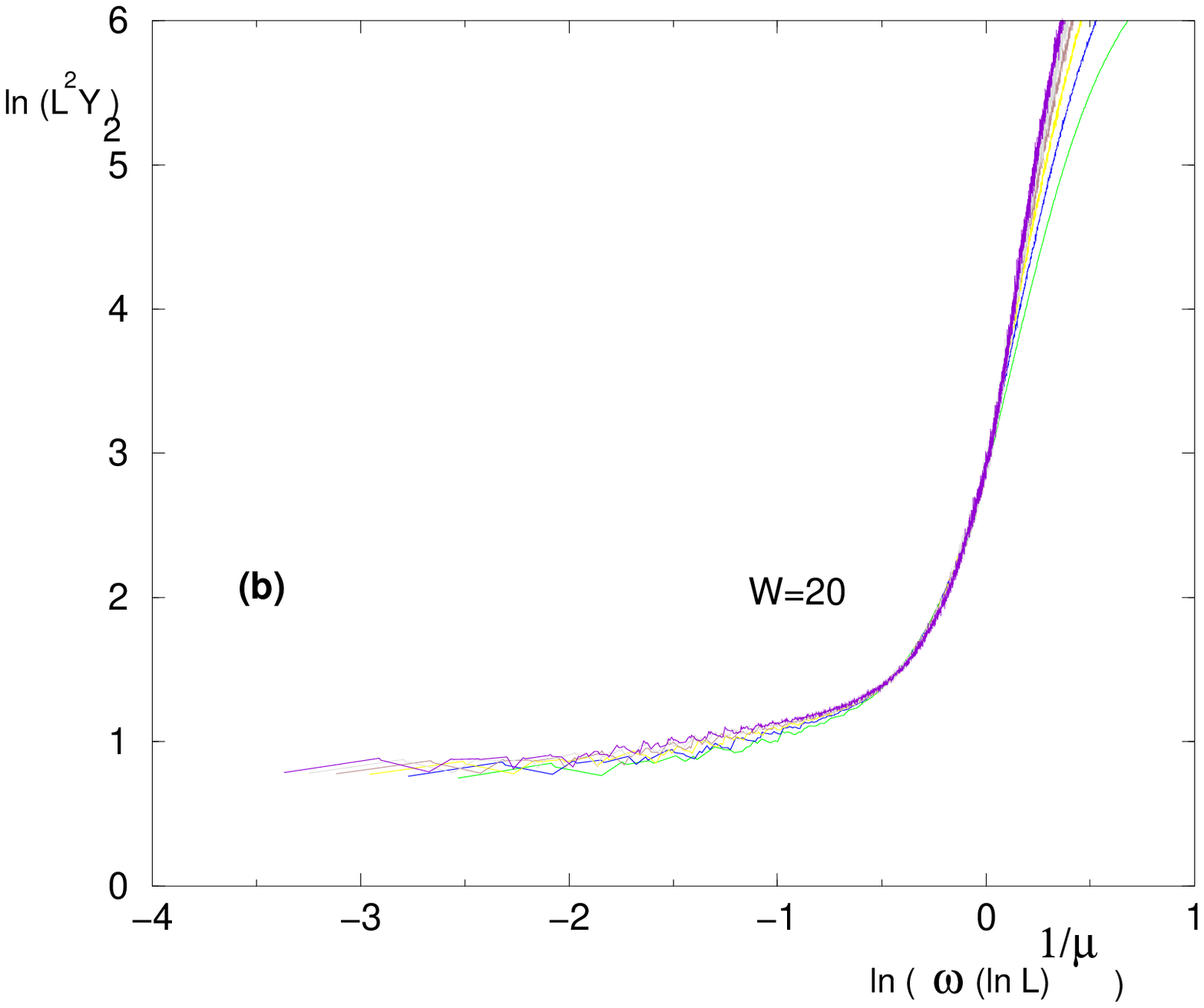}
\caption{(Color on line)    Finite-size scaling analysis of the low-frequency modes
in $d=2$ for $30 \leq L \leq 80$
 according to Eq. \ref{fssy2d} : 
 $y= L^2 Y^{typ}_2 (\omega,L)$ as a function of
 $z=(\ln L)^{1/\mu} \omega $ with the value $\mu=2$
(a)  data collapse in log-log coordinates for $W=1$
(b)  data collapse in log-log coordinates for $W=20$  }
\label{figfss2d}
\end{figure}

We show on Fig. \ref{figfss2d} that our data are compatible with the following 
finite-size scaling for the low-energy modes
\begin{eqnarray}
Y^{typ}_2 (\omega,L) \simeq \frac{1}{L^2} F_{d=2} \left( (\ln L)^{1/\mu} \omega \right)
 \ {\rm with }
\ \ \mu=2
\label{fssy2d}
\end{eqnarray}
This means that to each frequency $\omega$, one can associate a correlation length
$\xi(\omega)$ diverging as the following essential singularity
\begin{eqnarray}
\ln \xi(\omega) \oppropto_{\omega \to 0} \frac{1}{\omega^{2}}
\label{xid=2}
\end{eqnarray}
in agreement with \cite{john_Sompo}.
As in dimension $d=1$,
the conclusion is thus that any fixed frequency mode $\omega$ becomes localized in
the limit $L \to +\infty$  \cite{john_Sompo}.

However if one is interested into the set of eigenstates of a sample of a given size $L^2$,
the conclusion is that frequencies $\omega \ge (\ln L)^{-1/2}$ are localized,
whereas a certain number ${\cal N}_{deloc}(L)$ of eigenstates with frequencies
$\omega \le (\ln L)^{-1/2}$ are delocalized. 
From the behavior in $\omega$
of the integrated density of states of Eq. \ref{integrateddos2d}, one obtains that 
the fraction of delocalized states scales as the square of the pseudo-critical value 
$\omega^*(L) \sim  (\ln L)^{-1/2}$
 \begin{eqnarray}
\frac{{\cal N}_{deloc}(L)}{L^2 } \propto \left( \omega^*(L) \right)^2 \sim (\ln L)^{-1}
\label{fracndeloc2d}
\end{eqnarray}
So the number of delocalized eigenstates in a sample of size $L^2$ grows as 
 \begin{eqnarray}
{\cal N}_{deloc}(L) \oppropto_{L \to +\infty}  \frac{L^2}{\ln L}
\label{ndeloc2d}
\end{eqnarray}

\section{ Localization-delocalization transition of phonons in dimension $d=3$ } 

\label{dim3}

In this section, we present our numerical results obtained in dimension $d=3$
for the following sizes $L$ and the corresponding 
number $n_s(L)$ of disordered samples
\begin{eqnarray}
L && = 8,10, 12,14,16,18
\nonumber \\
n_s(L) && = 35.10^4,45.10^3,5500,1500,300,300
\label{nume3d}
\end{eqnarray}

\subsection{  Typical Inverse Participation Ratio $Y_2^{typ}(\omega,L)$ }

\begin{figure}[htbp]
 \includegraphics[height=6cm]{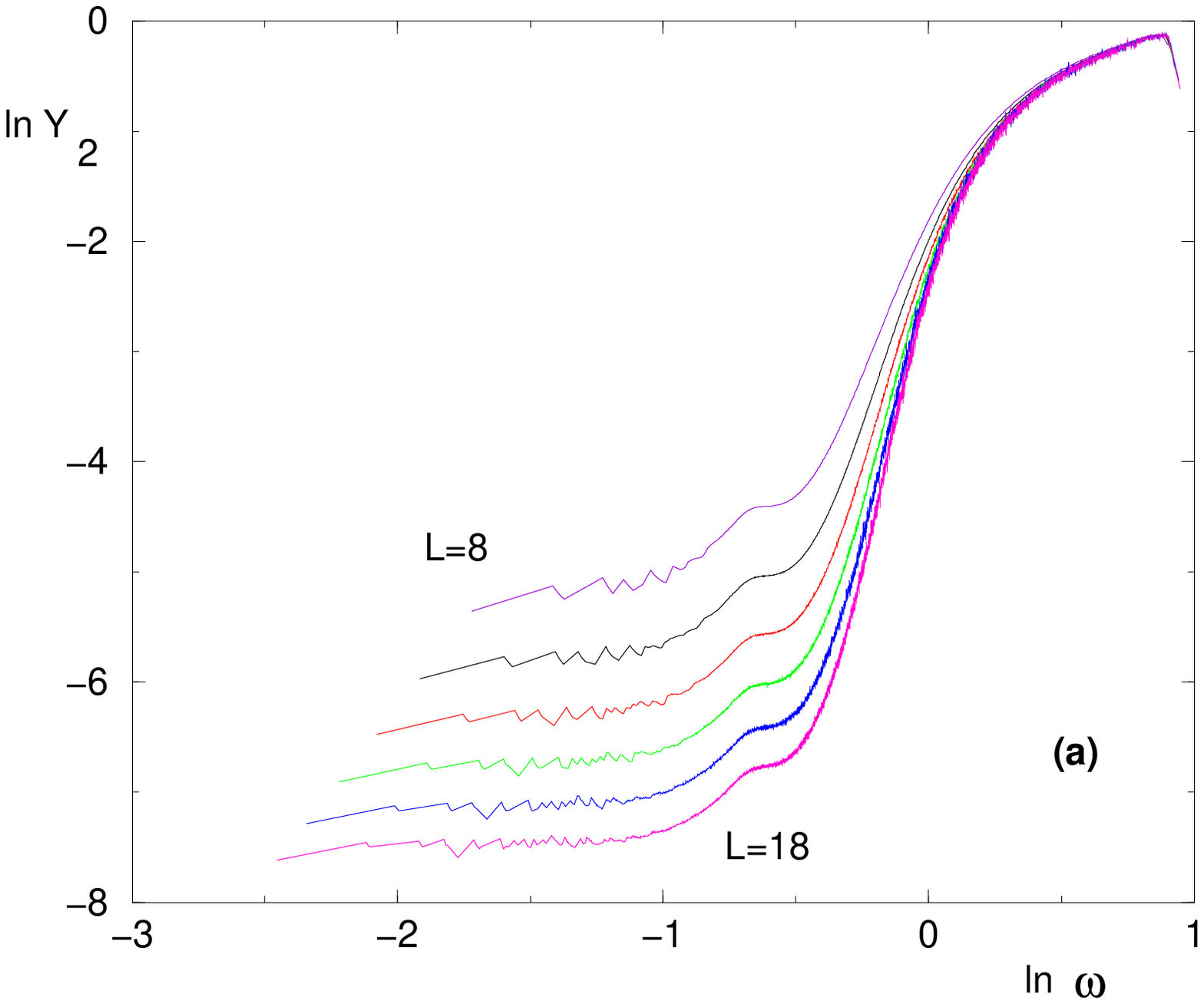}
\vspace{1cm}
 \includegraphics[height=6cm]{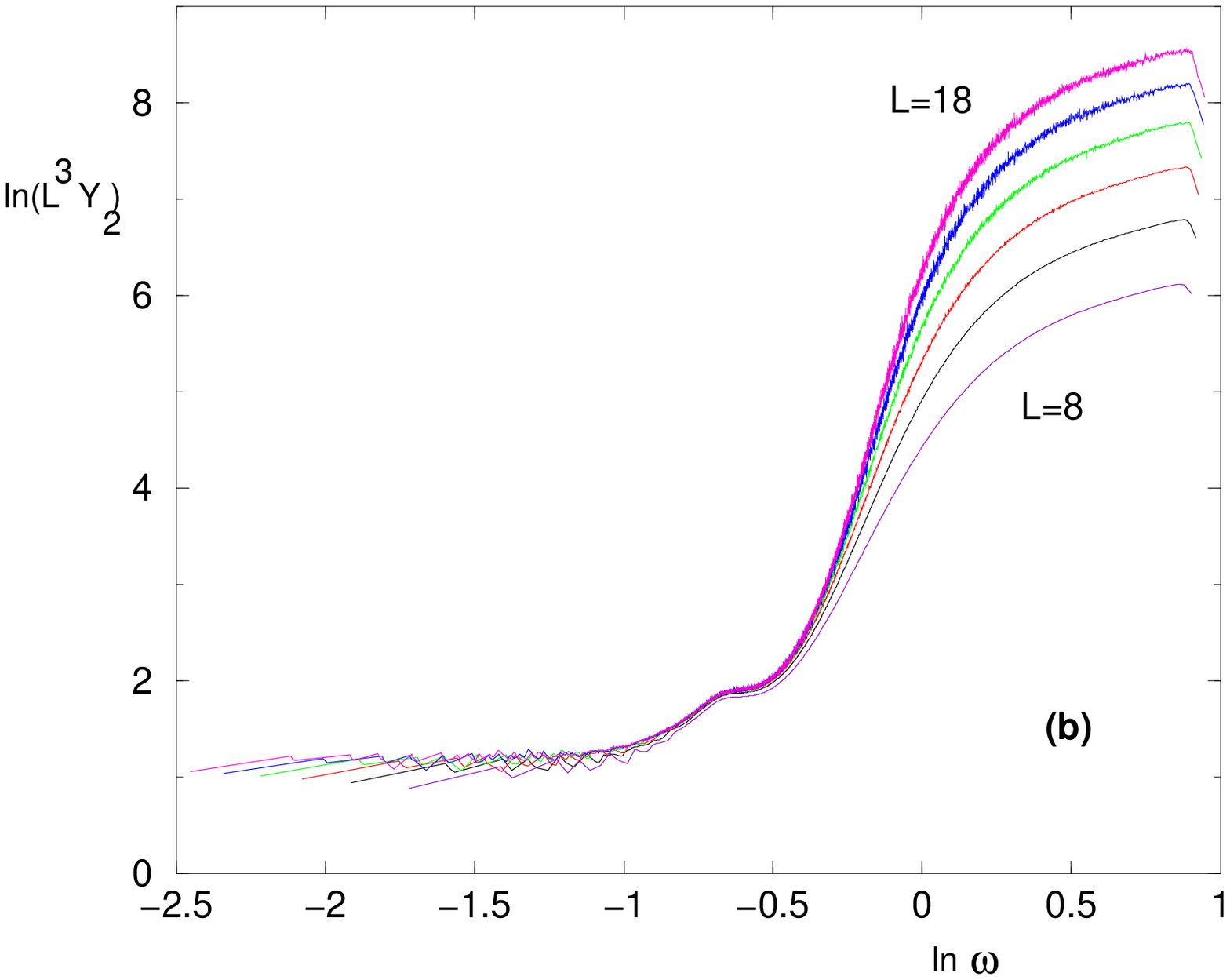}
\caption{(Color on line)  Typical IPR $Y_2^{typ}(\omega,L)$
as a function of the typical frequency in $d=3$ for
all eigenmodes of samples of sizes $8 \leq L \leq 18$ for $W=20$
(a) $\ln Y_2^{typ}(\omega,L)$ as a function of $\ln \omega$ : the collapse in the high frequency region corresponds to localized states 
(b) $\ln (L^3 Y_2^{typ}(\omega,L)) $ as a function of $\ln \omega$ : the collapse in the low frequency region corresponds to delocalized states 
 }
\label{figomipp3d}
\end{figure}

We show our data for the typical I.P.R. $Y_2^{typ}(\omega,L)$ on Fig. \ref{figomipp3d} (a) :
in the high-frequency part of the spectrum, the data collapse for the various sizes $L$
corresponds to localized states with a finite value $Y_2^{typ}(\omega,\infty)>0$
(see Eq. \ref{iprloc}). We show on Fig. \ref{figomipp3d} (b) the same data after
the appropriate rescaling $L^3 Y_2^{typ}(\omega,L)$ to detect the delocalized states (see
Eq. \ref{iprdeloc}) : the data collapse in the low-frequency part of the spectrum
corresponds to delocalized states.

\subsection{ Finite-size scaling analysis of the localization transition }

\begin{figure}[htbp]
 \includegraphics[height=6cm]{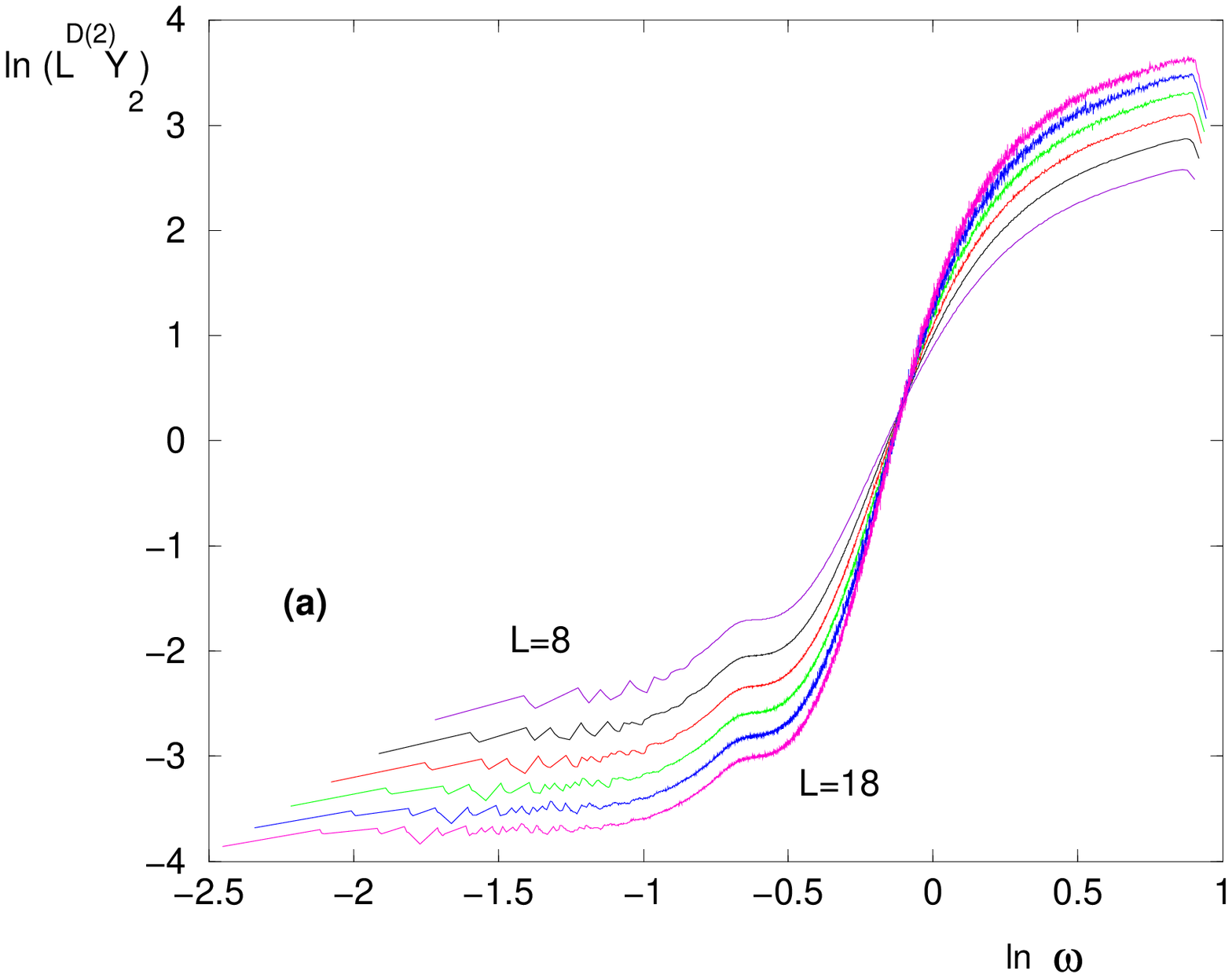}
\vspace{1cm}
 \includegraphics[height=6cm]{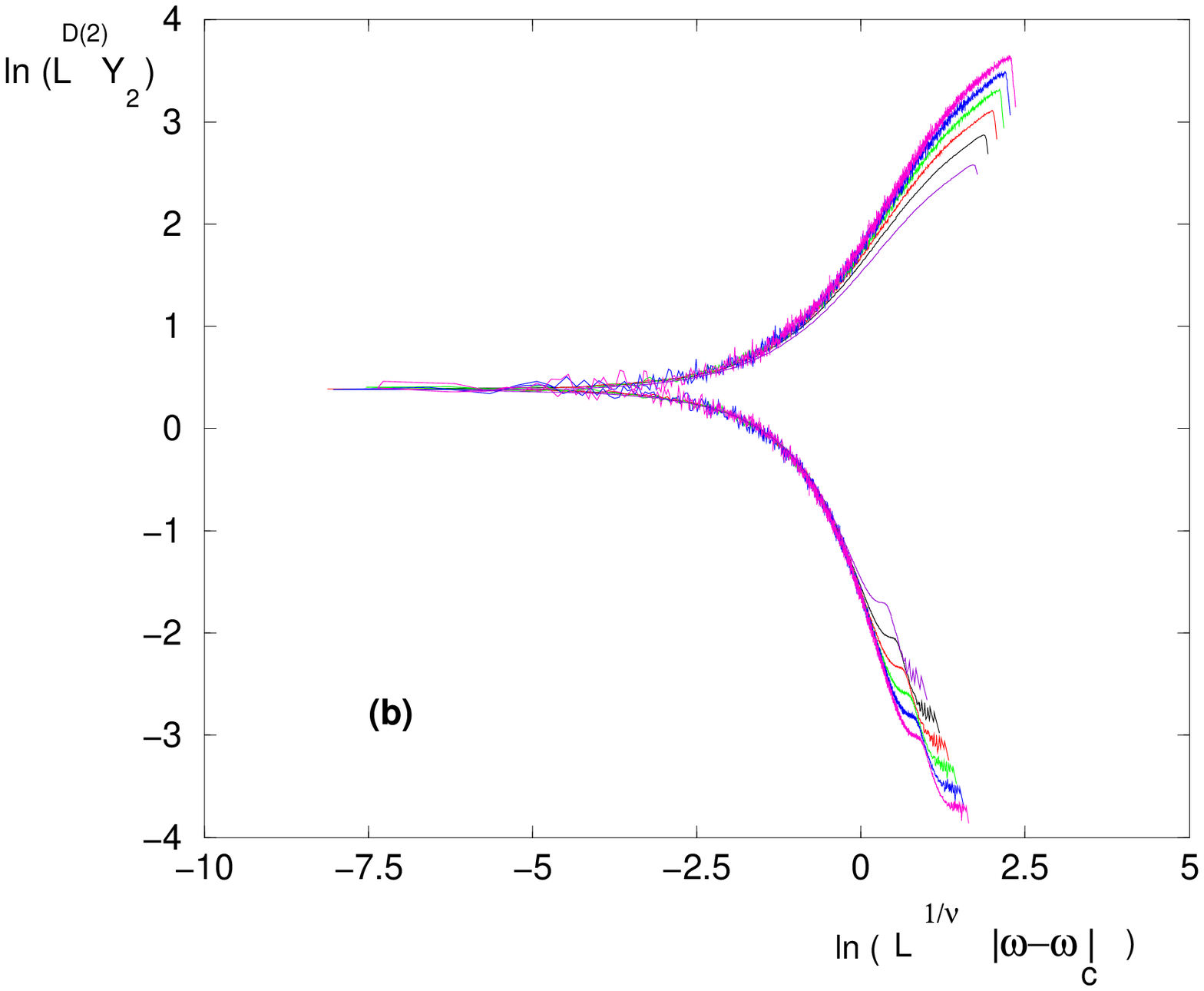}
\caption{(Color on line)  Finite-size scaling analysis of the localization transition
in $d=3$ for the disorder strength $W=20$
(a) $\ln (L^{D(2)} Y_2(\omega,L))$ as a function of $\ln \omega$ for the sizes
$8 \leq L \leq 18$ with the value $D(2) = 1.3$ : the crossing determines the critical point $\ln \omega_c(W=20) \simeq -0.1$
(b) $\ln (L^{D(2)} Y_2(\omega,L))$ as a function of $\ln (\omega-\omega_c)+(1/\nu) \ln L$ for $\nu =1.57$ : the data collapse is satisfactory. }
\label{figcriti3d}
\end{figure}

In dimension $d=3$, one expects that there exists a localization-delocalization transition
at some finite frequency $\omega_c>0$ \cite{john_Sompo}. The I.P.R. is then expected
to follow the following finite-size scaling 
\begin{eqnarray}
Y^{typ}_2 (\omega,L) \simeq \frac{1}{L^{D(2)}} F_{d=3} \left( L^{1/\nu} (\omega-\omega_c) \right)
\label{fssy2d=3}
\end{eqnarray}
The exponent $D(2)$ governs the power-law decay of the I.P.R. exactly at criticality
\begin{eqnarray}
Y^{typ}_2 (\omega_c,L) \propto \frac{1}{L^{D(2)}} 
\label{y2criti}
\end{eqnarray}
For the transition of the Anderson tight-binding electronic model in $d=3$,
it has been measured numerically (see \cite{mirlinrevue} and references therein)
\begin{eqnarray}
D_{Anderson}(2) \simeq 1.3
\label{d2}
\end{eqnarray}
As shown on Fig. \ref{figcriti3d} (a), 
if we rescale our data using this value, we obtain that 
the curves $L^{D(2)} Y^{typ}_2 (\omega,L)$ for various $L$ cross
  around the value $\ln \omega_c (W=20) \simeq -0.1$ corresponding to
\begin{eqnarray}
 \omega_c (W=20) \simeq 0.9
\label{omegac}
\end{eqnarray}
The integrated density of states at this value is around $N(\omega_c (W=20)) \simeq 0.66$ (data not shown), so that the critical point is sufficiently inside the spectrum to
have enough localized states and delocalized states on both sides
(this is not the case for any value of the disorder strength as explained below in section \ref{toosmallw}).

In addition, if we now rescale our data in terms of the reduced variable 
$L^{1/\nu} (\omega-\omega_c)$ with the value 
of the correlation exponent
that has been measured numerically 
for the Anderson tight-binding electronic model in $d=3$
(see \cite{mirlinrevue} and references therein)
\begin{eqnarray}
\nu_{Anderson} \simeq 1.57
\label{nu}
\end{eqnarray}
we obtain a good data collapse as shown on Fig. \ref{figcriti3d} (b).
Our conclusion is thus that the localization transition of phonons in $d=3$ is
governed by the same universality class as the 
Anderson tight-binding electronic model in $d=3$.

\subsection{ On the importance to consider strong enough disorder $W$ to observe the transition }

\label{toosmallw}

\begin{figure}[htbp]
 \includegraphics[height=6cm]{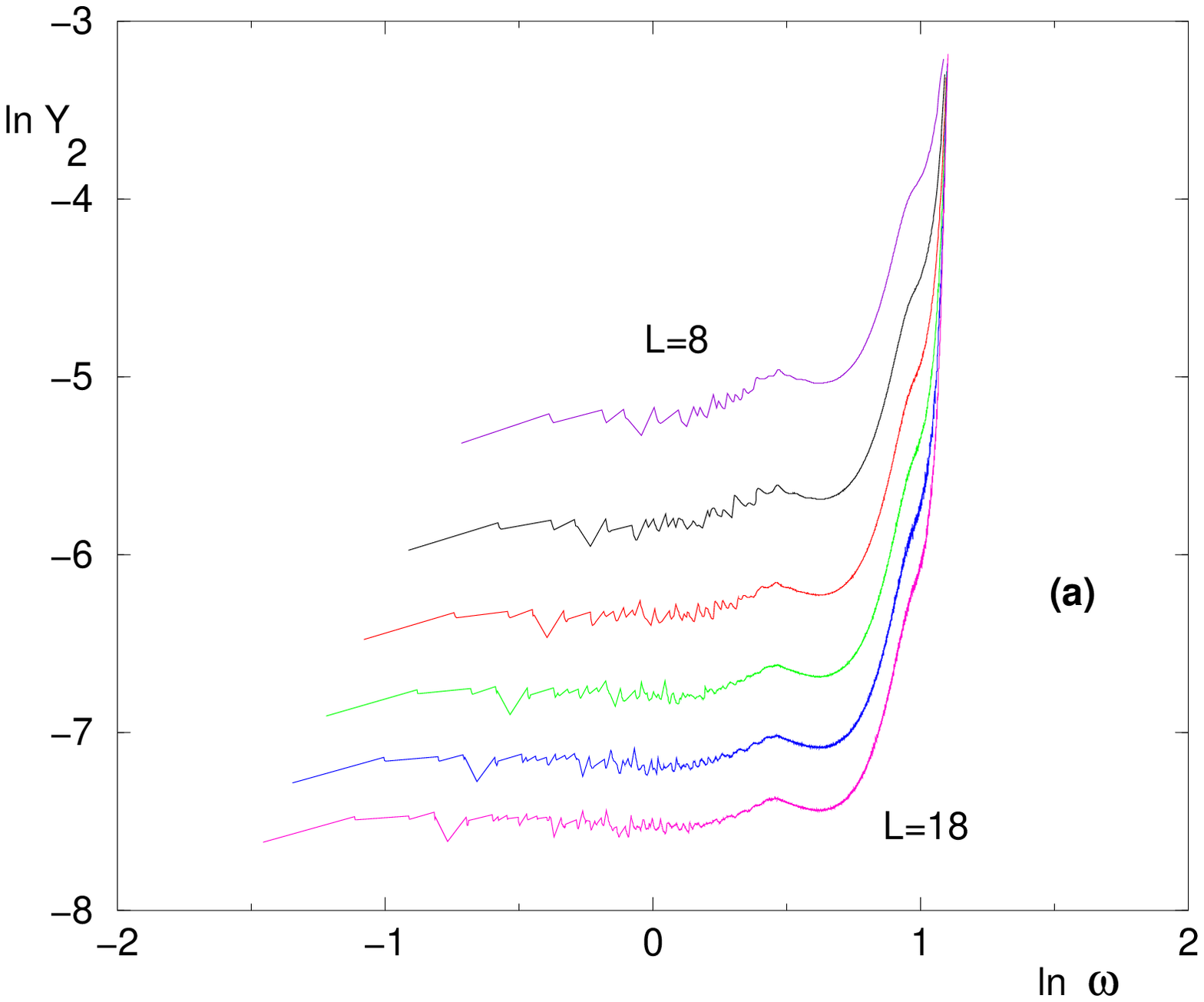}
\vspace{1cm}
 \includegraphics[height=6cm]{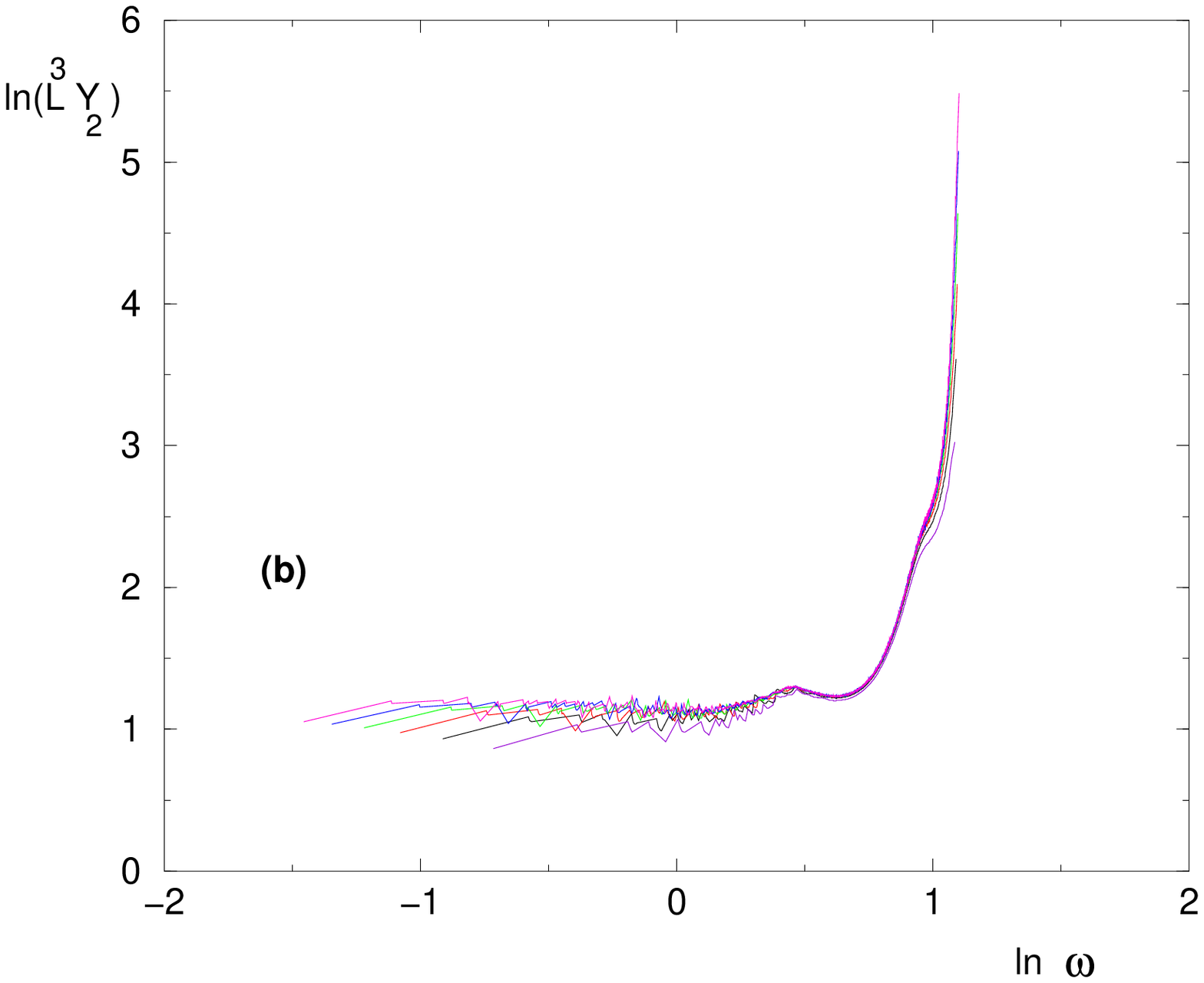}
\caption{(Color on line) 
Typical IPR as a function of the frequency $\omega$ in $d=3$ for
all eigenmodes of samples of sizes $8 \leq L \leq 18$ for the disorder strength $W=1$
(a) $ \ln Y_2^{typ}$ as a function of $ \ln \omega$ : even in the region of the highest frequencies, there is no collapse that would correspond to localized states 
(see the difference with the data corresponding to $W=20$ on Fig. \ref{figomipp3d} (a))
(b) $\ln (L^3 Y_2^{typ}) $ as a function of $\ln \omega$ : the collapse on the whole frequency region means that all states are delocalized 
(see the difference with the data corresponding to $W=20$ on Fig. \ref{figomipp3d} (b))
 }
\label{figomipp3dw=1}
\end{figure}

Up to now, we have described our results in dimension $d=3$ for the disorder strength
$W=20$, for which we have found a clear localization transition 
at some finite frequency $\omega_c(W=20)$ (Eq. \ref{omegac})
well inside the spectrum. However, as in dimensions $d=1$ and $d=2$, we have also
studied the disorder strength $W=1$ : the corresponding data for the
typical I.P.R. $ Y^{typ}_2 (\omega,L)$ shown on Fig. 
\ref{figomipp3dw=1} indicate that here all states of the spectrum are actually delocalized.
This does not mean that there is no critical frequency $\omega_c(W=1)$,
but that this critical value is not accessible, because it is higher that
the maximal frequency $\omega_{max}$ of the spectrum where
the density of states reaches its asymptotic value $N(\omega_{max})=1$.
Our conclusion is thus that to observe the localization transition, 
one should consider sufficiently strong disorder $W$ to ensure
that the corresponding critical value belongs to the spectrum
\begin{eqnarray}
\omega_c(W) < \omega_{max}(W)
\label{critereW}
\end{eqnarray}
We have found that this condition is satisfied for $W=20$, but is not satisfied for 
$W=1$. Changing the value of $W$ allows to move the critical value $\omega_c(W)$
inside the spectrum. 

\section{ Conclusion } 

\label{sec_conclusion}

To characterize the localization properties of eigenstates for phonons
in the presence of random masses in dimension $d=1,2,3$, we have studied numerically
the behavior of the typical Inverse Participation Ratio $Y_2(\omega,L)$ as a function of the frequency $\omega$ and of the linear length $L$ of the disordered samples. 

In dimensions $d=1$ and $d=2$, we have found that the low-frequency part $\omega \to 0$ of the spectrum satisfies the following finite-size scaling  
$L Y_2(\omega,L)=F_{d=1}(L^{1/2} \omega)$ in dimension $d=1$
and $L^2 Y_2(\omega,L)=F_{d=2}((\ln L)^{1/2} \omega)$ in dimension $d=2$.
We have moreover explained that the loose statement ``all eigenstates are localized
in dimensions $d=1,2$'' should be stated with some care for phonons :
 it is true that an eigenstate of any fixed frequency $\omega$ becomes localized in the limit $L \to +\infty$, but one should also be aware that
 a given disordered sample of fixed length $L$ contains 
a certain number $N_{deloc}(L)$ of delocalized states growing as $N_{deloc}(L)\sim L^{1/2}$
in $d=1$ and as $N_{deloc}(L)\sim L^2/(\ln L)$ in $d=2$.
These low-frequency delocalized modes are expected to play a major role in 
the dynamical properties on large distances, and in particular in
the heat transport problem where the disordered sample is connected to heat baths at the boundaries (see for instance \cite{Leb_spo}). 

In dimension $d=3$, for strong enough disorder strength $W$ ($W=20$ in our case), 
we have found a very clear 
localization-delocalization transition at some finite critical frequency
 $\omega_c(W)>0$. We have shown that our data are compatible 
with the finite-size scaling  $L^{D(2)} Y_2(\omega,L)=F_{d=3}(L^{1/\nu} (\omega-\omega_c))$
with the values $D(2) \simeq 1.3$ and $\nu \simeq 1.57$
 corresponding to the universality class
of the localization transition
 for the Anderson tight-binding electronic model in dimension $d=3$.
We have also found that for too small disorder strength (namely $W=1$ in our case)
the critical point $\omega_c(W)$ can be higher that the maximal frequency of the spectrum,
so that all eigenstates are actually delocalized.
The choice of too small disorder strengths seems to be the reason
why localized states were found only very near band-edges in previous numerical studies
\cite{akita,elliott,sep,Leb_spo} .

\end{document}